# Optimizing Data Extraction from Materials Science Literature: A Study of Tools Using Large Language Models

Wenkai Ning,[a] Musen Li,*[abc] Jeffrey R. Reimers,*[ac] and Rika Kobayashi*[d]

Large Language Models (LLMs) are increasingly utilized for large-scale extraction and organization of unstructured data owing to their exceptional Natural Language Processing (NLP) capabilities. Empowering materials design, vast amounts of data from experiments and simulations are scattered across numerous scientific publications, but high-quality experimental databases are scarce. This study considers the effectiveness and practicality of five representative AI tools (ChemDataExtractor, BERT-PSIE, ChatExtract, LangChain, and Kimi) to extract bandgaps from 200 randomly selected Materials Science publications in two presentations (arXiv and publisher versions), comparing the results to those obtained by human processing. Although the integrity of data extraction has not met expectations, encouraging results have been achieved in terms of precision and the ability to eliminate irrelevant papers from human consideration. Our analysis highlights both the strengths and limitations of these tools, offering insights into improving future data extraction techniques for enhanced scientific discovery and innovation. In conjunction with recent research, we provide guidance on feasible improvements for future data extraction methodologies, helping to bridge the gap between unstructured scientific data and structured, actionable databases.

## 1 Introduction

In recent years, the role of Large Language Models (LLMs) in scientific research has gained increasing attention, with a growing number of studies focusing on Artificial-Intelligence (AI)-driven methods for extracting data from scientific literature.[1, 2] LLMs have demonstrated exceptional capabilities in processing unstructured data, allowing researchers to efficiently extract and organize large volumes of information.[3-8] This capability is especially important in fields like Materials Science, where experimental data is often scattered throughout vast bodies of literature and presented in complex, inconsistent formats. Unlike computational data, which originates from computers and is typically structured and easy to manage,[9] experimental data requires manual identification and integration—an effort that is both time consuming and labour intensive. As a result, there is often a shortage of high-quality structured experimental data. Leveraging the power of LLMs, it should be possible to overcome these challenges and construct comprehensive databases that can significantly enhance data-driven discovery and accelerate scientific progress.[3, 10-13]

Currently, despite the promising applications of LLMs, there remains a significant challenge: the lack of a standardized, universally effective method for literature data extraction. Existing projects utilizing LLMs often employ varying approaches,[3, 10-19] each with distinct methodologies, making it difficult for researchers to determine the most effective strategies for extracting relevant and accurate data from unstructured sources.

The information extraction is a sub-task of Natural Language Processing (NLP). Traditional extraction ways, which do not necessarily involve LLMs, follow the NLP methods, for example, Named Entity Recognition (NER), Relation Classification (RC, or Relation Extraction), and Event Extraction (EE), etc.[20-23] NER can identify various parts of the text and find valuable elements (i.e. entities) in sentences (e.g., compounds, descriptions of material properties, numerical values, units), making it easier to extract the desired data. It is a crucial part of extraction and can be seen as a pre-task for RC, which identifies relationships between entities. The primary goal of RC is to accurately pair entities, especially when dealing with multiple similar types. Thus, an effective method involves creating a pipeline that integrates both NER and RC.[24, 25]

After the rise of LLMs as a powerful NLP tool, researchers began to use them for data extraction tasks. For example, Carlini et al. used GPT-2 to extract hundreds of verbatim sequences from training data.[26] Dunn et al. showcase fine-tuned GPT-3's ability to perform joint NER and RC for hierarchical information in Materials Science.[12] Foppiano et al. evaluated three LLMs (GPT-3.5-Turbo, GPT-4, and GPT-4-Turbo) for information extraction tasks in Materials Science, benchmarking them against BERT-based models and rule-based approaches. Their results revealed that LLMs excel at reasoning and relation extraction.[10] Dagdelen et al. fine-tuned LLMs (GPT-3 and Llama-2) for NER and relation extraction in Materials Science, showing that LLMs can accurately produce structured, hierarchical scientific knowledge from unstructured text.[3]

Using deep-learning architectures, LLMs are implemented using a transformer architecture that features an attention mechanism to enable better learning of semantic relationships.[27] There are two primary types of transformer, Bidirectional Encoder Representations from Transformers (BERT)[28] and Generative Pre-trained Transformer (GPT).[29] In terms of usage, BERT is typically used for classification tasks while GPT is used for generation tasks. Both approaches provide extraction options based on deep learning (Deep Learning-based NLP), corresponding to extraction based on NER (NER-based NLP).

Many extraction techniques have been developed, and here we discuss four main techniques through five of the most representative tools: ChemDataExtractor,[30, 31] BERT-PSIE,[14] ChatExtract,[15] LangChain,[32] and Kimi.[33] Key properties of these tools, and the relationships between them, are highlighted in Fig. 1.

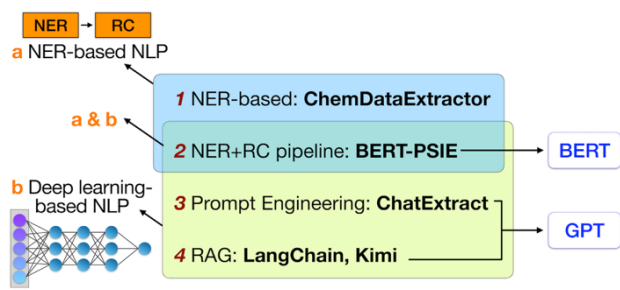

**Fig. 1.** Relation of the representative techniques and tools. ChemDataExtractor and BERT-PSIE, grouped within the blue rectangle, are based on NER. BERT-PSIE, ChatExtract, LangChain, and Kimi, grouped within the (overlapping) green rectangle, are based on deep learning. The second tool, BERT-PSIE, is a combination of both main NLP ways. In terms of transformer types, BERT-PSIE uses BERT models, while ChatExtract, LangChain, and Kimi use GPT models.

ChemDataExtractor[23] and its improved version, ChemDataExtractor 2.0,[24] are widely used tools in the field, they have some built-in modules for parsing sentences, and users need to define rules for the target property value. They can extract nested data (different data connected to each other) simultaneously if nested rules are defined, e.g. bandgap value at a specific temperature. They do NLP without using LLMs.

BERT-PSIE[14] (BERT Precise Scientific Information Extractor) consists of three different models to the three main parts of the workflow. First is to filter out the sentences that contain data from papers. Then NER is completed and the words labelled. Finally, RC finds the right match and outputs the structured data. This tool provides a typical NER plus RC pipeline, combined with BERT.

ChatExtract[15] implements a specific set of prompts designed to extract data from papers, which is known as Prompt Engineering, a process that involves refining and crafting instructions for LLMs to ensure they generate accurate and relevant outputs. The key to this approach, as opposed to directly asking for information step by step, is that it introduces additional input (informing the model that it may have made mistakes during the extraction process), allowing it to reconsider and correct previously extracted information, which improves accuracy.

LangChain[32] is a framework that employs Retrieval-Augmented Generation (RAG)[34] to reduce hallucinations in LLMs—a phenomenon where models generate information that is non-existent. Hallucinations provides misinformation that undermines reliability. To enhance the accuracy and trustworthiness of LLMs in information retrieval systems and NLP tasks, researchers are actively exploring methods to detect and mitigate hallucinations.[35-41] The RAG technique is formed with three main parts: retrieval (R), augmentation (A), and generation (G) to prioritise information provided by the user. RAG first splits the original document into many paragraphs (chunks) using appropriate strategies, a step known as "chunking". Then, using *Embedding Models*, it organises the chunks into vector representations in a vector database, selecting the most similar chunks matching the inputted retrieval queries. In the augmentation step, these chunks are utilized as context information, being concatenated with the pre-defined extraction prompt. Finally, using the reduced text compared to one without RAG, the LLMs specified as *Inference Models*, to generate more accurate results.

In addition to LangChain, another RAG-based tool, Kimi,[33] is also used. It is an online Chatbot implementation based upon closed-source LLM.

In summary, this study explores the use of tools based on four representative NLP techniques (NER-based, NER+RC pipeline, Prompt Engineering, and RAG) to extract data from randomly-selected unstructured scientific publications. We evaluate these tools for performance with respect to human-extracted data for the extraction of bandgap data from scientific publications in Materials Science. In addition, we compare results for the LLMs (the four Deep Learning-based tools) with a traditional tool used for literature data extraction, ChemDataExtractor. A crucial element of the analysis is employing consistent evaluation criteria, enabling an objective comparison of the effectiveness of the different tools.

By doing so, we provide insights that can help researchers select the most suitable tool for their specific projects. Furthermore, we identify areas for improvement in existing methodologies, with the long-term goal of developing a more generalizable approach that can serve the needs of Materials Science and other fields reliant on structured data extraction from scientific literature. This work presents a comprehensive overview of the data extraction methods, serving as a guide for future research.

**Table 1.** Table of data structure, showing four examples of *Source* information and how this is represented in terms of *Material* name and measurement conditions, bandgap *Value*, *Position Class* (see Table 2) and *Value Class* (see Table 3). For the complete table, please refer to the Excel file on manual data extraction in ESI.

| Paper | Position Class | Value Class | Material | Value (eV) | Source |
|---|---|---|---|---|---|
| Ref.[42] | 1 | 1 | SrFBiS2 | 0.8 | … SrFBiS2 is a semiconductor with a direct bandgap of **0.8 eV** … |
| Ref.[43] | 2 | 2 | H-MoSe2 | ~ 1.13 | H-MoSe2 … with a direct bandgap of **about 1.13 eV** while T-MoSe2, ZT-MoSe2 and SO-MoSe2 are zero bandgap materials. |
| Ref.[43] | 2, 4 | 1 | T-MoSe2 | 0 | H-MoSe2 … with a direct bandgap of about 1.13 eV while T-MoSe2, ZT-MoSe2 and SO-MoSe2 are **zero bandgap** materials; also in Table 1 of the original paper. |
| Ref.[44] | 3 | 2 | SmN (AFM phase, Optical absorption measurements) | ~ 0.7 | … were limited to the AFM phase of SmN. …(*sentences*)… Optical absorption measurements indicate the existence of a gap of **about 0.7 eV**. |

## 2 Methods

The source code, inputs and outputs, and user instructions needed to reproduce our data extractions are provided in full on GitHub.[45] Except where explicitly noted, all results should be reproducible from this information. New applications of these tools may readily be developed using this resource.

### 2.1 Collection of the dataset

To make unbiased datasets, we collected metadata of arXiv papers using the dataset from Kaggle.[46] Then we filtered out papers in the Materials Science field (in which "categories" contains "mtrl-sci", the identifier for Materials Science in arXiv). To balance the quality and quantity of the dataset (minimizing scanned PDF files as much as possible while maximizing the number of articles), the publication time span was restricted from January 1, 2000, to November 1, 2024. This process selected 93,391 papers. From these, 200 papers were selected randomly to form the evaluation set of this work, and both the arXiv pre-print and publisher's versions were downloaded to make two, naively equivalent, datasets.

In total, 37 of the 200 published papers and 39[42-44, 47-82] of their arXiv variants were identified by human means as containing bandgap data. Including papers identified by one or more AI methods as also containing such data, these counts increase to 149 of the publisher papers and 142 of their arXiv variants. Hence the issue of the bandgap is seen to be very relevant in Materials Science, enabling our two diverse and unbiased datasets to expose manifold aspects of the data extraction process.

### 2.2 Classification of data presentation

Drawing upon our experience in reading scientific literature and manually extracting data, we identified diverse ways in which material data was presented in articles. This analysis led us to categorize these data presentation formats from various perspectives. To illustrate this, Table 1 shows four examples of bandgap property information found by human analysis, listing the identifiers: *Position Class*, *Value Class*, *Material*, *Value*, and *Source*. The two classes differentiate data presentations for subsequent analysis. The *Material* variable includes the materials' name, along with any categorical property required to uniquely identify the reported bandgap, *Value* depicts the bandgap value in eV, and *Source* refers to the original sentence(s) from which the data were taken, which facilitates subsequent checking of the AI-extracted results.

Based on the location of the data, we divide the data into the five *Position Classes* listed in Table 2: "Single Mention", "Multiple Mention", "Context", "Table", and "Figure". Single Mention indicates that both two main pieces of information for a material, *Material* and *Value*, appear in the same sentence just once, as in the examples from Ref.[42] listed in Table 1. Alternatively, Multiple Mention indicates that either *Materials* and/or *Values* are mentioned more than once, e.g., the two extracted records from Ref.[43] listed in Table 1.

**Table 2.** *Position Classes* describing the textual relationships between *Material* description and bandgap *Value*, and the number of human-extracted records obtained from the 200 publisher-version[a] papers considered.

| Pos. Class | Description | Count[b] |
|---|---|---|
| 1 | Single Mention: A *Material* and its property *Value* are mentioned once and in the same sentence. | 43 |
| 2 | Multiple Mention: More than one *Material* and property *Value*s are mentioned in the same sentence. | 96 |
| 3 | Context: Features identifying/categorising *Material* and *Value* are in different sentences. This may overlap with other categories. | 70 |
| 4 | Table: Data appears in the tables of the paper. | 82 |
| 5 | Figure: Data appears in the figures of the paper. | N/A[c] |

a: See "comparison" spreadsheets in ESI for details of results for both the publisher and arXiv versions.

b: Up to two relationships have been ascribed to each data record, see ESI.

c: Data in figures is not considered in this work.

Context represents instances where *Material* and *Value* appear across different sentences, or the name of the material requires contextual information to be fully represented; this is quite common since materials are often replaced by pronouns, requiring analysing of the prior text which explains what that pronoun refers to. This type of data challenges a tool's ability to read and summarize context.

The last two categories, Table and Figure, refer to cases where the material property data is in tables or figures of papers. For the processing of tables, since the tools we use involve first converting them into text and then extracting information from that text, the extraction effect varies depending on the tool's ability to understand the processed text. Indeed, only LangChain and Kimi can extract data from tables and, in this work, data in figures are not considered. In principle, there are tools that would facilitate extracting data from figures, including those that recover the original data behind charts[83, 84] and those that "understand" images by aligning image pixels and text descriptions.[85, 86] Nevertheless, to date, such tools have not achieved the robustness needed to handle arbitrary images layouts, making quantitative analysis premature.

In addition, according to the type of numerical values present, we categorise the data into five *Value Classes* as listed in Table 3: "Fixed", "Estimated", "Range", "Bounded", and "Change". The ideal situation is an unambiguous single Fixed value, for example in Table 1 from Ref.[42], "SrFBiS2 is a semiconductor with a direct bandgap of 0.8 eV". However, sometimes the data is not a Fixed value and appears as in Table 1 from Ref.[44] as "a gap of about 0.7 eV", which falls into the second category, Estimated value. A common alternate form of this is "~ 0.7 eV". In addition, sentences like "with a bandgap more than 5 eV" are categorized as Bounded. We classify sentences similar to "the bandgap decreases from 2.07 eV by 0.15 eV" as Change, which adds considerable complexity to the data.

**Table 3.** *Value Classes* describing the type of bandgap *Value*, and the number of human-extracted records obtained from the 200 publisher-version[a] papers considered.

| Value Class | Description | Count[b] |
|---|---|---|
| 1 | Fixed (e.g., = 1.07 eV) | 144 |
| 2 | Estimated (e.g., ~ 1.07 eV) | 28 |
| 3 | Range (e.g., ± 0.02 eV) | 47 |
| 4 | Bounded (e.g., < 1.07 eV) | 3 |
| 5 | Change (e.g., increased by 0.3 eV) | 3 |

a: See "comparison" spreadsheets in ESI for details of results for both the publisher and arXiv versions.
b: Up to two classes have been ascribed to each data record, see ESI.

### 2.3 Data Processing

In this work, we used papers in PDF format as input because all the papers within the selected time span contain PDF versions. PDFs without a fixed structure are more difficult to parse than formats such as HTML/XML, so subsequent extension to including other formats should be straightforward.

Herein, we uniformly used PyMuPDF[87] to parse PDF files and converted them into plain text, processing them into one sentence per line and saving them to TXT files. The subsequent processing of the five tools varied. For ChemDataExtractor and ChatExtract, the TXT file obtained from the previous step was directly inputted, whereas ChatExtract required the sentences presented line by line to meet the requirement for sentence-by-sentence processing. For BERT-PSIE, we further processed the TXT file into JSON format[88] (a lightweight data interchange format easy for humans to read and for machines to parse) to meet its input requirements. LangChain required unprocessed PDF files to fit the mechanism of the original code, but PyMuPDF[87] was still used for parsing. For Kimi, we directly imported the original PDF file without doing any data processing.

After extraction, post-processing was performed. First, we unified the outputs in different formats, which yielded the initial extracted data. Then, we conducted a preliminary cleaning, removing entries that are, for example, duplicated or clearly not our target data, resulting in cleaned data. Finally, we normalized the data, unifying the units to eV and formats of representation, for final evaluation and comparison. Of note, optical bandgaps reported as wavelengths were not included in any of the analyses presented.

### 2.4 Methodological implementations

Our application of ChemDataExtractor referenced the method developed by Dong et al.[16] for extracting bandgap and temperature values, using their Snowball model applied to the unified evaluation dataset we prepared. Owing to compatibility issues with the package, this tool was implemented indirectly through use of Docker,[89] which unfortunately was computationally very inefficient.

The original BERT-PSIE project provided models trained to extract bandgap values and Curie temperature values.[14] We modified the code to apply its three fine-tuned models to only extract bandgap values.

One of the focuses of ChatExtract's original work[15] was to design a group of extraction prompts, providing different prompts based on the feedback from the *Inference Model* used. We further conducted Prompt Engineering based on the original work's prompt group to adapt to our unified post-processing, detailed prompts provided in ESI. In addition, we modified the online GPT-4 model that they chose, which is related to our Kimi implementation (see ESI Section S2), to one of three widely used offline open-source *Inference Models*: Llama2:13b,[90] Llama3.1:70b,[91] and Qwen2.5:14b,[92] see Table 4. All three models were encapsulated in GGUF format to balance model size and performance.[93] GGUF is a binary format that is designed for fast loading and saving of models, as well as ease of reading.

**Table 4.** *Embedding Models* and *Inference Models* used by different variants of ChatExtract (CE) and LangChain (LC).

| Name | Embedding Model | Inference Model |
|---|---|---|
| $CE_1$ | N/A | Llama2:13b |
| $CE_2$ | N/A | Llama3.1:70b |
| $CE_3$ | N/A | Qwen2.5:14b |
| $LC_{11}$ | Nomic-embed-text | Llama2:13b |
| $LC_{12}$ | Nomic-embed-text | Llama3.1:70b |
| $LC_{13}$ | Nomic-embed-text | Qwen2.5:14b |
| $LC_{21}$ | Bge-m3 | Llama2:13b |
| $LC_{22}$ | Bge-m3 | Llama3.1:70b |
| $LC_{23}$ | Bge-m3 | Qwen2.5:14b |

The three locally deployed *Inference Models* for ChatExtract were also utilized in LangChain's generation phase (G, the third phase in RAG). Before that, two *Embedding Models* were implemented in the first phase, retrieval (R), to process the plain text data produced from the PDF files: Nomic-embed-text[94] and Bge-m3.[95]

As an alternative to LangChain, we also tested a different RAG tool, Kimi-k1.5,[33] which is a closed-source online model known for its ability to read long texts (see ESI Section S2 for details on how it implements RAG). We extracted data by manually uploading PDF files one by one to its official webpage and using the prompt consistent with LangChain.

Some tools involve internal hyperparameters that can be adjusted to enhance performance. We used the hyperparameters specified in the original works if provided, and for LangChain we conducted optimisation tests on the hyperparameters (see Results section).

Regarding the deployment details of RAG, we used a separate vector database for each paper, rather than incorporating all PDF files into a single vector database for retrieval. Specifically, for LangChain, we processed one paper at a time, then cleaned the vector database and established a new vector database for the next paper; for Kimi, we opened a new chat window each time.

### 2.5 Performance criteria

Each extracted record from a paper, as well as the absence of extracted records, can be judged according to the following criteria. Firstly, "P" and "N" are used to represent "Positive" and "Negative" results respectively, judging whether a record is extracted. Secondly, "T" and "F" are used to represent "True" and "False" results respectively, judging whether the extraction situation is correct. Therefore TP, which stands for "True Positive", indicates the desired outcome of correctly extracted records, whereas TN, which stands for "True Negative", indicates the other desired outcome that no data is extracted from papers that do not contain data. Alternatively, FN indicates missed records that were not extracted (the label F here means that the unextracted situation is not ideal). Finally, FP stands for "False Positive" and indicates that wrong records were extracted, either for non-existent data (hallucinations) or else just incorrectly extracted (the label F here means that the extracted situation is not ideal). The relationships between TP, FP, FN, and TN are illustrated in Fig. 2; the number of extracted records of each type are named *TP*, *FP*, *FN*, and *TN*, respectively.

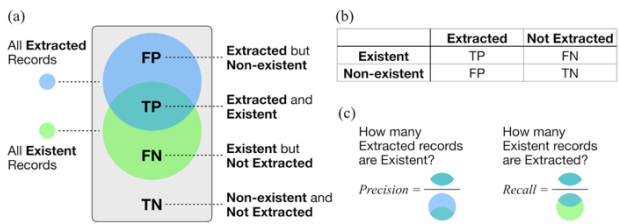

**Fig. 2.** Illustration of record classification based on extraction and existence status. (a) Venn diagram showing the relationship between all extracted records (blue circle) and all existing records (green circle). (b) Contingency table summarizing the classification outcomes for records based on their existence and extraction status.

The actual records contained within a paper were determined by human data extraction. This is a time-consuming and error-prone process and was conducted by two researchers in parallel. Differences between this manually extracted data and data from AI extraction were examined in detail, occasionally resulting in revision of the human-extracted data.

### 2.6 Classification of error types

To help understand the factors that lead to FP generation, seven *Error Classes* were identified and are listed in Table 5. Taking the most critical entity set—*Material*, *Value*, and *Unit*—as an example, if there is a complete record including these three items, and the automatically extracted data correctly captures *Value* and *Unit*, but misses or incorrectly extracts *Material*, the result is classified as FP. If the data exists but the extracted *Value* is incorrect, then this is also described as FP. Hallucinations are examples of FP results in which data is extracted whereas no data was actually present. For more details of the strict matching mechanisms used and case analyses, please refer to ESI Section S3. Subtle details pertaining to the error class analysis are also provided in ESI Section S4.

**Table 5.** *Error Classes* contributing to FP results, and the number of materials for which each class was identified from the 200 publisher-version[a] papers considered.

| Error Class | Description | Count[b] |
|---|---|---|
| 1 | Hallucination: Extracted *Material* or *Value* absent in paper. | 265 |
| 2 | Insufficient *Material*: The description of *Material* is inadequate, missing necessary categorical information. | 294 |
| 3 | Wrong *Material*: *Material* exists but no related bandgap in the paper. | 394 |
| 4 | Inaccurate *Value*: Wrong Value Class, e.g., missing "~" in "~1.45 eV". | 114 |
| 5 | Wrong *Value*: *Value* exists but is not a bandgap. | 461 |
| 6 | Wrong *Unit*: Error extracting *Unit*, e.g., "100 meV" vs. "100 eV". | 40 |
| 7 | Wrong pairing: Two (or more) *Materials* have their *Values* interchanged. | 14 |

a: See "comparison" spreadsheets in ESI for details of results for both the publisher and arXiv versions.
b: Up to two classes have been ascribed to each data record, see ESI Section S4.

### 2.7 Performance metrics

Using the *TP*, *TN*, *FP*, and *FP* counts from our 200-examined papers in each dataset, we calculate the established *Precision*, *Recall*, and *F-score* metrics:[96-98]

$$Precision = \frac{TP}{TP + FP} = \frac{\text{correctly extracted}}{\text{all extracted}} \quad (1)$$

$$Recall = \frac{TP}{TP + FN} = \frac{\text{correctly extracted}}{\text{all existent}} \quad (2)$$

$$F\text{-}score = 2 \cdot \frac{Precision \cdot Recall}{Precision + Recall} \quad (3)$$

and introduce a new metric:

$$Null\text{-}Precision = \frac{TN}{null\ papers} = \frac{correctly\ eliminated\ papers}{all\ null\text{-}result\ papers} \quad (4)$$

For all four metrics, the higher the score the better the performance.

*Precision* is the fraction of correctly extracted records amongst all extracted ones, whereas *Recall* is the fraction of correctly extracted records from amongst all the actual records. Simply speaking, *Precision* focuses on the accuracy of the model, whereas *Recall* focuses on the integrity (completeness) of the generated results database. If a model has high *Precision* but low *Recall*, it means the database is accurate but incomplete. Conversely, a model with high *Recall* but low *Precision* is guessing—it is complete, but its contents are of low accuracy. The *F-score* is the harmonic mean of *Precision* and *Recall*, balancing the two.

*Null-Precision* is the ability of the tool to exclude papers that are of no value, simplifying subsequent (human) processing of extracted data. Unlike the data entries for the other three metrics, *Null-Precision* is calculated at the article level. It utilizes TN, which is not considered in the other metrics. If the *Null-Precision* is low, then many papers containing only FPs will be brought to attention, a highly undesirable result. If *Null-Precision* is high but *Precision* low, then the FPs are concentrated in certain papers and absent from others, an improved result; ideally both *Null-Precision* and *Precision* should be high. Combined, metrics provides a comprehensive perspective for analysis of the extraction results.

Many groups have considered how to interpret analysis metrics.[10, 14, 15, 30-32, 99-102] There are, however, no generally accepted values for what constitutes "good" and "bad" results, and not all metrics always need to be high; details depends on the purpose of the study. For an exploratory study seeking information on a new topic, perhaps simply a high *Null-Precision* (> 90 %) would suffice. Alternatively, a study could desire high *Precision* (> 90 %), when simply a set of examples with accurate data is required. Then again, one may wish to know all possibilities, for which high *Recall* (> 90 %) is required. To build a large database that is too extensive for human curation, then a high *F-score* (> 90 %) is required.

## 2.8 Statistical analysis

The 200 papers were processed as a whole to determine the performance metrics. To estimate the likely errors arising from the small sample size, the papers were divided into four sets of 50, and the metric standard deviations used to estimate error bars:

$$error\ bar = \frac{standard\ devation}{\sqrt{N}} \quad (5)$$

where $N = 4$ is the number of datasets.

## 3 Results

The records initially extracted, cleaned, and then normalised by the 12 tool variants are provided on GitHub, with the human extracted data provided in Excel spreadsheets in ESI, for both the publisher-version and arXiv-version datasets. The normalised data are then combined into final "comparison" Excel files available in ESI and statistically analysed (results that appear in the main text Tables, Figures, and Text are coloured therein in red). Of note, except for the optimisation of the explicitly described hyperparameters, no data from the comparison spreadsheets was used to influence the tools or Python code that initially processed the 200 papers.

### 3.1 Unexpected differences found between the arXiv and publisher datasets

Naively, it was expected that few differences would arise between the results obtained from analysing the arXiv and publisher versions of the papers, except in the uncommon circumstance that pertinent changes were made during the publication process. Instead, significant differences were found, with Fig. 3 highlighting the changes in the *F-scores* obtained from extractions performed by each of the AI tools.

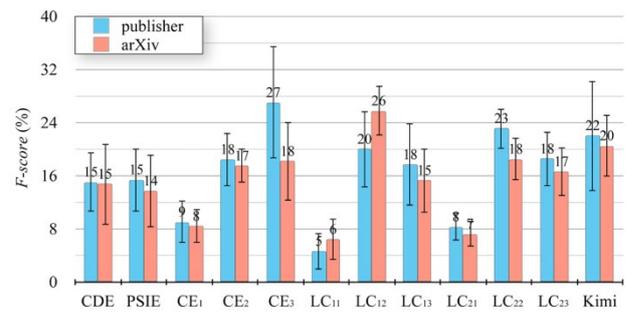

**Fig. 3.** Bar graph showing the *F-scores*, with error bars, obtained by the 12 AI tools from 200 arXiv and publisher version papers.

The *F-scores* from the publisher PDFs were found to be slightly higher than those from arXiv PDFs for most tools, with differences ranging from marginal (e.g., CDE, 15 % vs. 15 %) to more pronounced (e.g., $CE_3$, 27 % vs. 18 %). A notable exception is $LC_{12}$, for which the arXiv version yielded a higher *F-score* (26 %) compared to the publisher version (20 %).

Most publisher versions embody complex and variable typesetting structures and additional information outside of the main text, whereas the structures of the arXiv versions are typically simpler and more uniform, see ESI for details on the differences in the presentation of paper content. This could account for the slightly greater error bars obtained for most tools.

Of particular interest are the results from $CE_3$, $LC_{12}$, and $LC_{22}$ for which the error bars obtained by dividing each dataset into four subsets are similar to the substantial differences found between the arXiv and publisher results, see Fig. 3. For example, for $CE_3$, the error bars are ± 6 % for arXiv and ± 8 % for publisher, with the difference between sources being 9 %. In this case, including the arXiv and publisher versions together, as if they were *independent* publications, to make four subsets of 100 papers, yields an error bar remaining at ± 6 %. This shows that the randomness found *between* different data subsets can match that found when analysing two versions of the *same* paper. Therefore, these two different effects could have the same cause.

Human comparison of the text generated after parsing the example PDF file shows that the text processed by each tool between the arXiv and the publisher version is mostly small local changes. This was found for both the arXiv and publisher versions, see ESI. Hence the differences do indeed arise from the presentation of the local text in different global contexts, be that arising from author-generated presentation variations between papers or else from different presentations of the same paper.

### 3.2 Understanding RAG context analysis

To better understand the sensitivity of the observed extracted results to data presentation, we considered the mechanism of RAG in the information extraction process. We took one example paper[50] and fed different sized fractions of its content to the RAG analysis, with the size ranging from the bare minimum containing just the single feature of interest to the whole text.

Starting from the 179th sentence, which contains the content "ZnO band gap (3.4 eV)", we began with 10 sentences centred around it, expanding five sentences before and after each time, ultimately expanding to the entire article. This generated a total of 37 sets of text with gradually increasing volume for further processing. We gave them one by one to all six LangChain variants for complete processing and additionally recorded the output of each step of RAG, with the analysis summarized in the ESI.

During processing, the text is divided into chunks, which are then fed into the RAG retrieval phase. In the retrieval phase, the *Embedding Model* calculates the similarity between each chunk and the retrieval query (see Introduction Section on RAG), and then ranks them. The results presented in ESI Section S6 show that, the retrieval query has a significant impact on the extraction results. As the amount of text in one of the 37 processed sets increases, the resulting changes in chunk segmentation alter the structure of the vector database. Even if the data-containing chunks are identical, the *Embedding Model* used in the retrieval phase can give different vector similarity scores, owing to the different global contexts in which the chunks are located. Consequently, newly added irrelevant text may obtain higher similarity scores than the target chunk, potentially causing the loss of target data and its subsequent recovery as more text is incorporated. This is one of the reasons for the difference in scores between arXiv and publisher versions of the same article and indicates intrinsic weaknesses in the RAG methods. As a means of optimising and controlling data extraction, RAG provides hyperparameters that could, in principle, alleviate these observed effects.

### 3.3 RAG hyperparameter selection

The most significant hyperparameters in the RAG approach are believed to be: *Temperature*,[103] *Chunk-size*, *Chunk-overlap*, and *Top-k*.[104] Hyperparameters are adjustable parameters used to define any configurable part of the learning and inference processes of LLMs. Unlike the fixed parameters of LLMs after training, the hyperparameter settings used during inference can affect the output. We sought to improve the results by optimizing these hyperparameters.

**3.3.1 Optimising the LangChain *Temperature*.** *Temperature* is one of the most important hyperparameters in LLMs, applied during the inference stage to rescale the model's *logits* (i.e., the raw output scores before softmax normalization), controlling the randomness of the predicted token distribution. In LLMs, a *token* represents the smallest unit of text the model can interpret and produce, typically corresponding to a word, subword, or symbol. It serves as the fundamental processing element on which the model operates during both training and generation.

When *Temperature* = 0, the output is deterministic and reproducible, but otherwise a random element appears so that replicated extractions from the same paper yield different results. According to Li et al.,[103] non-zero *Temperatures* may improve results, but there is no single *Temperature* established that can adapt to all LLMs across all types of tasks. We accordingly investigated the performance of multiple *Temperature* values in the extraction task.

We conducted extraction experiments on the 200 publisher versions using different *Temperatures* for all six variants of LangChain, and the resulting *F-scores* are shown in Fig. 4. The *Temperature* values used were 0, 0.25, and 0.8, with the set at 0.8 repeated a second time to judge the reproducibility of extractions with non-zero temperature. The figure shows that adjusting *Temperature* does not have a significant effect. Hence, as the deterministic output at a *Temperature* value of 0 is beneficial for reproducibility and does not affecting the quality of the extraction results, we choose this hyperparameter for subsequent extraction work. This hyperparameter is also applied to the LLMs of ChatExtract.

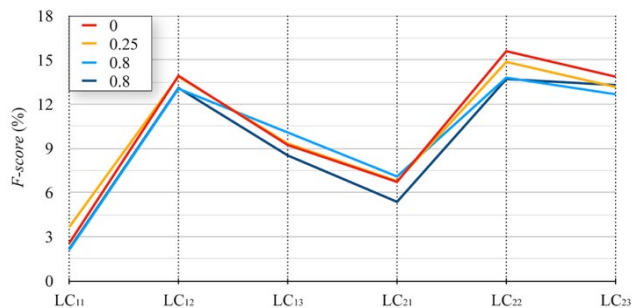

**Fig. 4.** The effect of varying *Temperature* on the *F-score* achieved by the six LangChain variants, setting values of 0, 0.25, 0.8, and then a repeated extraction at 0.8.

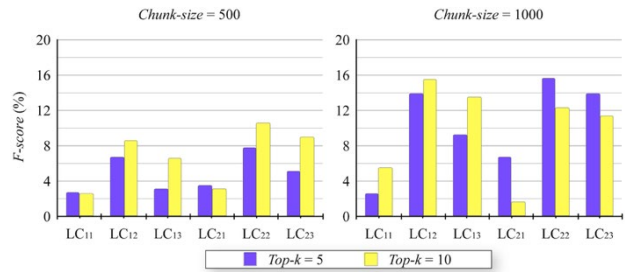

**Fig. 5.** Bar graph presents a *F-score* comprehensive analysis of how the *Top-k* and *Chunk-size* hyperparameters independently influence performance for the six LangChain methods.

**3.3.2 Optimising *Chunk-size* and *Top-k* of LangChain.** The hyperparameters *Chunk-size* and *Chunk-overlap* control how text is segmented and overlapped before being embedded into the vector database. The former determines the maximum number of tokens allowed when splitting the original text into chunks, whereas the latter controls how many tokens can overlap between consecutive chunks. This acts to prevent useful information from being inadvertently split, separating say the *Material* and *Value* identifiers. In this study, *Chunk-overlap* is set to 1/5 of *Chunk-size*, leaving only one adjustable hyperparameter.

The final hyperparameter, *Top-k*, determines the number of priority chunks retrieved during the retrieval step of RAG. *Chunk-size* and *Top-k* operate in the first two phases of RAG, collectively determining the text supplied to the LLMs for prediction in the third phase. We employ recursive chunking to hierarchically divide text into smaller segments until each meets the token limit, so *Chunk-size* is slightly larger than the actual number of tokens in each chunk. Therefore, the amount of text seen by the LLMs can be roughly estimated as the product of *Chunk-size* and *Top-k*.

We conducted four sets of experiments by varying *Chunk-size* and *Top-k*; detailed descriptions are provided in the ESI. Figure 5 compares the effects of *Top-k* and *Chunk-size*, with data presented as *F-scores*. The comparison shows that *Chunk-size* = 1000 generally achieves higher scores than *Chunk-size* = 500. When *Chunk-size* = 500, increasing *Top-k* consistently improves results, seemingly suggesting that more contextual information is better. However, when *Chunk-size* = 1000, increasing *Top-k* actually decreases *F-scores* for $LC_{21}$, $LC_{22}$, and $LC_{23}$, indicating that this assumption is not valid, the optimal context amount is between the two states, and we choose the overall better *Chunk-size* = 1000 and *Top-k* = 10 for subsequent extractions. Overall, *Chunk-size* plays the dominant role, with *Top-k* providing opportunities for fine-tuning in specific-use cases.

### 3.4 Performance of the 12 AI tools

Henceforth, analysis is performed only for the optimised choices of hyperparameters and for the publisher versions of the 200 papers. Human analysis revealed *TP+FN* = 220 as the number of *existent* records. This number is used to determine *Recall*, see Eqn. (2). Notably, these 220 data records come from 37 of the 200 papers, which means that the *number of null papers* = 163; this value is used to determine *Null-Precision*, see Eqn. (4).

Table 6 summarizes the performance of 12 AI tool variants. It lists the total number of AI-extracted results (*TP+FP*), as well as *TP*, *Precision*, *Recall*, *F-score* and *Null-Precision*, obtained using tools based on ChemDataExtractor,[30, 31] BERT-PSIE,[14] ChatExtract,[15] LangChain,[32] and Kimi[33]. The results for *Precision*, *Recall*, *F-score*, and *Null-Precision* are also presented graphically in Fig. 6(a), along with estimated error bars (see comparison spreadsheets in ESI for full details). The error bars are sufficiently small to indicate the robustness of the major qualitative features identified, but large enough to require caution in making detailed quantitative comparisons.

**Table 6.** Evaluation of extracted results (Eqns. 1-4) from the ChemDataExtractor[30, 31] (CDE), BERT-PSIE[14] (PSIE), ChatExtract[15] (CE), LangChain[32] (LC), and Kimi-k1.5[33] (Kimi), including the relative computational time cost.

| Tool | TP+FP [a] | TP | Precision (%) | Recall (%) | F-score (%) | Null-Precision (%) | Rel. Cost |
|---|---|---|---|---|---|---|---|
| CDE | 47 | 20 | 43 | 9 | 15 | 100 | 185[b] |
| PSIE | 54 | 21 | 39 | 10 | 15 | 99 | 1 |
| $CE_1$ | 424 | 29 | 7 | 13 | 9 | 58 | 102 |
| $CE_2$ | 52 | 25 | 48 | 11 | 18 | 98 | 252 |
| $CE_3$ | 99 | 43 | 43 | 20 | 27 | 97 | 65 |
| $LC_{11}$ | 174 | 9 | 5 | 4 | 5 | 67 | 4 |
| $LC_{12}$ | 51 | 27 | 53 | 12 | 20 | 98 | 17 |
| $LC_{13}$ | 64 | 25 | 39 | 11 | 18 | 96 | 4 |
| $LC_{21}$ | 168 | 16 | 10 | 7 | 8 | 64 | 5 |
| $LC_{22}$ | 75 | 34 | 45 | 15 | 23 | 94 | 18 |
| $LC_{23}$ | 83 | 28 | 34 | 13 | 18 | 94 | 5 |
| Kimi | 153 | 41 | 27 | 19 | 22 | 99 | N/A |

a: The sum *TP+FP* is the total count of data records extracted by each tool variant (after necessary post-processing cleaning).

b: Software translation is required due to device compatibility, resulting in the high cost; CDE is normally considered to be a cheap tool.

Considering all of the results, the best performance was found for the Prompt Engineering-based ChatExtract tool $CE_3$. Compared to the 220 human-extracted records, it extracted $TP$ = 43 correct records, achieving a *Recall* of 20 % (Eqn. (2)), as well as $FP$ = 56 incorrect records, achieving a *Precision* of 43 % (Eqn. (1)) and therefore an *F-score* of 27 % (Eqn. (3)). Its *Null-Precision* is 97 % (Eqn. (4)), making errors only five times when considering the 163 papers that did not actually contain bandgap data.

We can clearly see from Table 6 that the three tool variants $CE_1$, $LC_{11}$, and $LC_{21}$ (which all use *Inference Model* Llama2:13b) performed poorly. They generated significantly lower *Precision* and *Null-Precision*, as well as excessive FPs (Fig. 6(b)) compared to the other tools. Therefore Llama2:13b is ineffective for bandgap extraction and is neglected in parts of the subsequent analysis.

With the exception of those using Llama2:13b, all tools showed *Null-Precisions* of at least 94 %, a generally useful result. For Kimi and $LC_{23}$, *Precision* was intermediary at 27 % and 34 %, respectively. The remaining tools delivered higher *Precision* in the range of 39–53 %, with error bars between ± 6 % and ± 12 %, and so these tools are not strongly differentiated using our sample set containing only 200 publications. Nevertheless, for *Precision*, the best results were obtained by $LC_{12}$ (53 ± 12 %) and $CE_2$ (48 ± 10 %). The results for *Recall* were generally poor, with the best results being for $CE_3$ at 20 ± 7 % and Kimi at 19 ± 7 %. This effect led to poor *F-score* results, with the apparent best results being $CE_3$ at 27 ± 8 %, $LC_{22}$ at 23 ± 3 %, and Kimi at 22 ± 8 %.

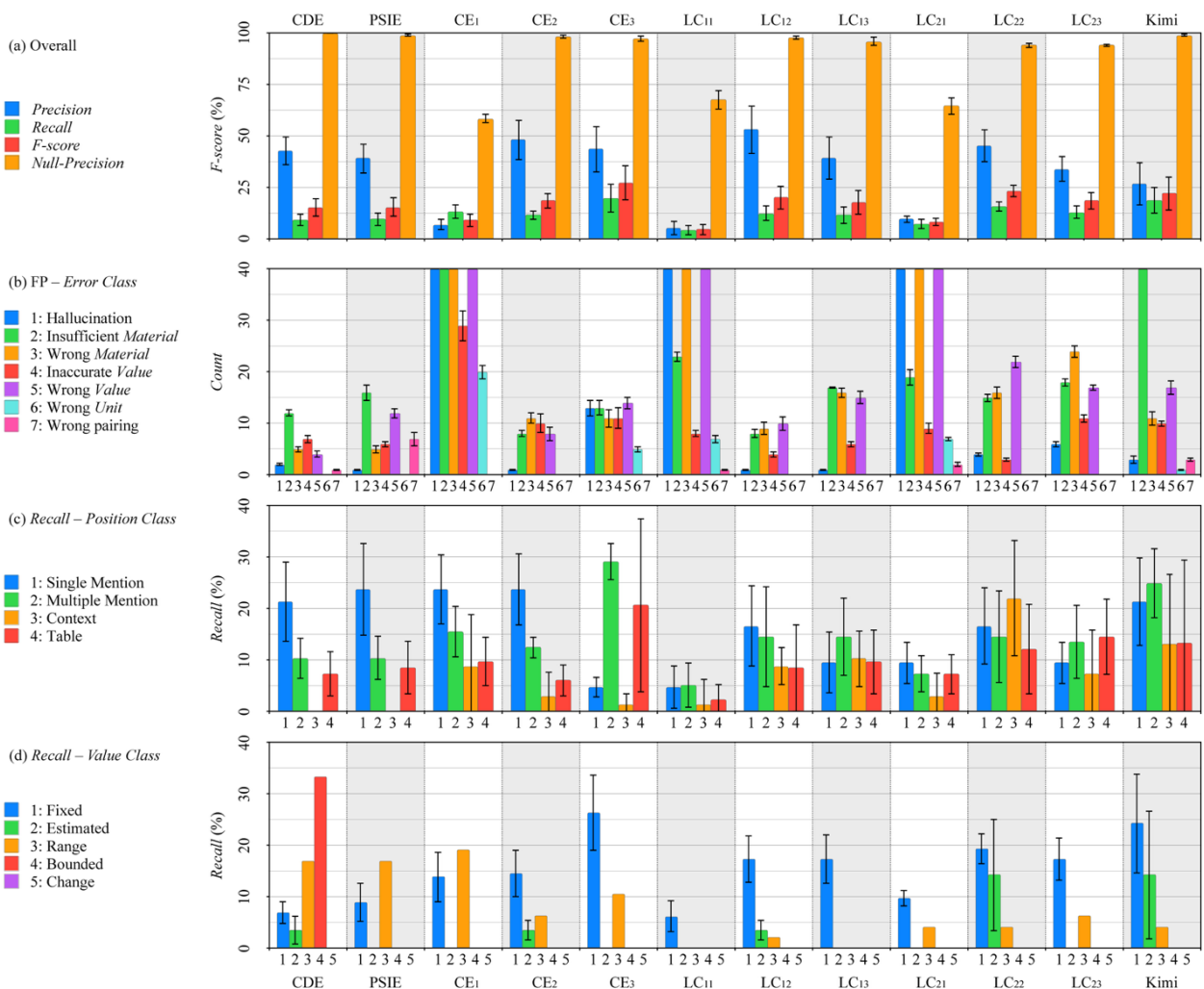

**Fig. 6**. Multi-perspective analysis of 12 tool variants using the publisher-version dataset (see ESI for data for the arXiv version). Error bars come from the analysis of four datasets of 50 papers each. The four subplots have a consistent arrangement of tools on the *x*-axis, i.e., the top and bottom of a block are the same tool. (a) Bar graph illustrating *Precision*, *Recall*, *F-score*, and *Null-Precision* metrics (see Table 6 for detailed data). (b) *Count* of *FP* categorized by *Error Class* (refer to Table 5); for *Counts* between 40 and 196, see ESI. (c) Bar graph depicting the relationship between *Recall* and *Position Class* (see Table 2); Class 5 not included. (d) Bar graph depicting the relationship between *Recall* and *Position Class*.

Taking data pertaining to models as a whole, a comparison between ChatExtract and LangChain variants with different end number shows that the *Inference Model* Qwen2.5:14b has better alignment with the questioning style of ChatExtract ($CE_3$ scores higher than $CE_2$ overall, and the *F-score* is much higher). We find that the extraction performance of $LC_{x2}$ is better than that for $LC_{x3}$, indicating that Llama3.1:70b suits LangChain better. Regarding the *Embedding Models*, a comparison between $LC_{1x}$ and $LC_{2x}$ shows that Bge-m3 results are generally better than those for Nomic-embed-text.

It is difficult to discriminate between the tool types (Fig. 1) owing to the magnitude of the calculated error bars, but the results for the NER-based tools, CDE and PSIE with *F-scores* of 15% each, excluding Llama2:13b, do appear to be less than those involving deep-learning NLP: CE (18–27 %), LC (18–23 %) and Kimi 22 %. It is *Recall* rather than *Precision* that most differentiates performance amongst these tool categories.

### 3.5 Correlation of false positive results with *Error Class*

As a guide to the mechanisms of failure of the 12 AI tool variants, Fig. 6(b) shows the number of extractions for each *Error Class* (see Table 5, which also lists the total number of errors in each class). The calculated error bars are shown in the figure and are relatively small compared to magnitude of the data variations.

The *Inference Model* found previously to work very poorly, Llama2:13b, consistently generated hallucinations and fails to identify either the *Material* or *Value* correctly. For the other tools, the failure to extract categorical variables needed for the unique identification of the *Material* is a recurring issue, but errors in *Units* and in pairing *Materials* with *Values* are uncommon. It appears that the tools just struggle to get all the details of the *Material* and its *Value* simultaneously correct. The diverse nature of the identified errors indicates that not one single NLP issue controls performance, with instead wide-ranging tool improvements being currently needed. From amongst the overall better-performing tools, Kimi makes most FPs (112), failing mostly to adequately characterise the *Material*. The tools with the least number of FP results are $LC_{12}$ (24) and $CE_2$ (27) (plus also CDE (27)), so a strong *Inference Model* (in this case Llama3.1:70b) helps to reduce errors.

### 3.6 Correlation of *Recall* with *Position Class*

The assignment of the *Position Class* (Table 2) is done by human means on the extracted records and therefore is only available for TP and FN results, leading to *Recall* (Eqn. 2). Figure 6(c) graphically depicts *Recall* for each *Position Class*, visualizing the capabilities of each tool variant. Class 5 has been excluded, because data extraction from images is not considered herein. As the subset of data considered is small, the error bars shown in the figure are sizable, yet small enough to identify key features.

For Class 3 (Context), Fig. 6(c), indicates *zero Recall* for CDE and PSIE, indicating that these tools completely lack contextual capability. Low *Recall* for contextual information is also found for CE, which uses sentence-by-sentence questioning. Contextual recognition is better facilitated by the RAG tools, with LC and Kimi delivering significantly better than those for other tool variants.

It is noteworthy that for even the seemingly simplest *Position Class* 1, in which the properties to be extracted appear alone in a single sentence, none of the tools achieved a *Recall* above 25%. After examining the *Error Classes* of the extracted data entries related to *Position Class* 1 (43 entries see Table 2), it was found that typically the *Error Class* was 2 (Wrong *Material*), 3 (Inaccurate *Material*), or 5 (Wrong *Value*), indicating the data was found but not accurately extracted for a variety of reasons.

Also, the best tool for extracting data present as *Position Class* 2 (Multiple Mention) and 4 (Table) was found to be $CE_3$. This feature contributes to it being the tool with the highest overall *Recall*.

Note that tools such as $CE_1$ that do not parse tables achieved non-zero scores for *Recall* from Tables as some of the data in the tables is also mentioned in the text.

### 3.7 Correlation of *Recall* with *Value Class*

Like *Position Class*, *Value Class* is only available for TP and FN results and hence is a descriptor of *Recall* only (Eqn. 2). Table 3 lists the number of records extracted from the publisher versions in each *Value Class*, with 144 having Class 1 (Fixed), 28 having Class 2 (Estimated) and 47 having Class 3 (Range). Figure 6(d) graphically depicts *Recall* for each tool, with the small amount of available data sometimes preventing error bars from being estimated or else resulting in large error ranges. *Recall* is highest for Fixed, as would be anticipated. Only $LC_{22}$ and Kimi produced comparable results for Estimated. The best results for Range were produced by CDE, PSIE, and $CE_1$. Only CDE correctly extracted results for Bounded, and no tool correctly extracted results for Change.

## 4 Discussion

### 4.1 Comprehensive evaluation

A summary of the advantages and disadvantages of ChemDataExtractor, BERT-PSIE, ChatExtract, LangChain, and Kimi, considering aspects of both tool implementation and results reliability, is provided in Table 7. This is presented under headings indicating performance, training, and ease of use, and combines aspects from the Results Section with practical aspects from the Method implementations.

From the performance perspective, we primarily consider features that govern the model's accuracy (*Precision*), extraction integrity (*Recall*), computational efficiency (see Table 6), and suitability as a starting point for further refinement. From the training perspective, we compare whether the model requires additional training or is ready to use, and whether it needs a large amount of human resources. Finally, we also consider user friendliness, which includes aspects like operational complexity and ease of getting started.

In summary, ChemDataExtractor is favoured by not needing training, and its ease of use, but its performance in this work is not ideal. Additionally, owing to the lack of maintenance of the software package, there are compatibility issues in deployment on some devices, and the indirect approach taken herein to its implementation led to large unnecessary computational costs. BERT-PSIE similarly has not demonstrated the expected level here, with both it and ChemDataExtractor being too cautious, only extracting very small amounts of data. ChatExtract does not need training, but it is very time-consuming owing to the large number of input and output of sentences. Of note, the tool CE$_3$ delivered the best *F-score* of all tools considered. The overall effect of LangChain is acceptable, and it does not need training, but the *Inference Model*, *Embedding Model* and associated hyperparameters need to be optimised. Most, but not all, LangChain variants delivered on the promise of avoiding hallucinations, with some ChatExtract tools showing similar performance. Kimi's performance is comparable with the best other tools considered and is easy to use, but the LLM's effects depend on the vendor, and the interactions depend on the network, limiting reproducibility.

**Table 7**. Evaluation of the practicality of the tools considered in terms of positive (+) and negative (−) features.

| Tool | Performance Accuracy, Integrity, and Efficiency | Training Difficulty and Manual Work | Ease of Use |
|---|---|---|---|
| CDE | + Includes table parser<br>+ Handles various expressions (pronouns)<br>+ Extracts nested properties<br>+ Excellent irrelevant content filtering<br>+ Can extract data ranges<br>+ Moderate *Precision*<br>− Undesirable *Recall*<br>− Cannot extract data in context<br>− Cannot extract from figures | + No training required<br>+ Single input, no additional operation required for extraction<br>− Requires parser training for optimization<br>− Needs repeated rule modifications for nested rules | + Built-in modules<br>+ Simple post-processing (structured output)<br>+ Extracts with sentence source for verification<br>− New rule definitions required for new properties<br>− Domain knowledge-dependent |
| PSIE | + Stable output<br>+ Can extract data ranges<br>+ Moderate *Precision*<br>− Undesirable *Recall*<br>− Cannot extract nested properties<br>− Poor context sensitivity (sentence-level)<br>− Cannot extract data in context<br>− Cannot extract from tables/figures | + Low training data requirement for new properties (fine-tuning)<br>+ Single input, no additional operation required for extraction<br>− Requires manual labelling of training data<br>− Challenging RC linear layer design | + No intricate grammar rules<br>+ Extracts with sentence source for verification<br>− Needs new models for new properties |
| CE | + Includes result confirmation<br>+ Extracts nested properties<br>+ Can extract data ranges<br>+ Best at extracting data from tables<br>+ Can show moderate *Precision*<br>− Poor context sensitivity (sentence-level)<br>− Undesirable *Recall*<br>− Cannot extract from figures<br>− Slow processing | + No training required<br>+ Single input, no additional operation required for extraction<br>− Needs testing for best LLMs | + Prompt modification suffices for new tasks<br>+ Extracts with sentence source for verification<br>− May require closed-source LLMs for optimal results<br>− Needs prompt adjustment for stability |
| LC | + Document-level processing<br>+ Extracts from tables/figures<br>+ Extracts nested properties<br>+ Context awareness (low effectiveness)<br>+ Can extract estimated values<br>+ Good match between cost effectiveness and performance<br>+ Can show moderate *Precision*<br>− Undesirable *Recall* | + No training required<br>+ Single input, no additional operation required for extraction<br>− Requires *Embedding Model* training (for better performance)<br>− Needs testing for best LLMs<br>− Contains hyperparameters that should be optimised | + Prompt modification suffices for new tasks<br>+ Deployment flexibility and scalability<br>− Extracting sentence sources simultaneously may result in a loss of accuracy |
| Kimi | + Document-level processing<br>+ Extracts from tables/figures<br>+ Extracts nested properties<br>+ Context awareness (low effectiveness)<br>+ Moderate *Precision* and *Recall*<br>− Accuracy depends on Online LLMs | + No training required<br>− Uncontrollable performance<br>− Model hyperparameters cannot be customized<br>− Requires per-article web interaction manually | + Prompt modification suffices for new tasks<br>+ Ready-to-use<br>− Extracting sentence sources simultaneously may result in a loss of accuracy<br>− Network-dependent |

For all models, parsing context information was a challenge. This could be the context of the *Material* and *Value* entities as used in the *Position Class*, but also generating the *Material* name from contextual information including the deciphering of pronouns. An example of this is provided from ref.[44] in Table 1. ChemDataExtractor and LangChain have made attempts at solving contextual issues, but the performance for the 200 papers considered is poor. ChemDataExtractor has a Forward-looking Dependency Resolution mechanism to deal with variations in naming. For example, white phosphorus is referred to as "white-P" throughout an article,[54] and the appropriate name translation must be made in the output records. In addition, it will also consolidate all data related to the same *Material* at the end. LangChain does this by accepting the entire paper into a vector database, only considering relevant paragraphs. Alternatively, both BERT-PSIE and ChatExtract process papers only at the sentence level and hence lack the ability to integrate contextual information.

Also, beyond the scope of this work, it is likely that the user will need to change the target of the information extraction tool from bandgap to other material properties. For example, in rule-based tools such as ChemDataExtractor, it is necessary to re-define the extraction rules, primarily by modifying the parsing expressions. To achieve comprehensive extraction, these parsing expressions must account for all potential variations, which requires both domain knowledge and iterative testing for refinement. In contrast, BERT-PSIE does not rely on predefined rules but instead requires training new models for each new property. This model training process demands manually labelled data, which can be time-consuming. However, it may not be necessary to retrain the model for each new property if the underlying BERT model is sufficiently robust. For ChatExtract and LangChain, the process is simplified as they only require specifying the name of the new property to the LLM, making migration straightforward.

### 4.2 Further prospects

Each tool has its own challenges. Currently, there is no tool that can achieve a level suitable for large-scale automated use. One possible recommendation for researchers regarding tool selection is that it is still necessary to use a combination of various tools and integrate the results. For example, we believe that Prompt Engineering (ChatExtract) can be used in conjunction with RAG technology, leveraging the ChatExtract's ability to reconsider and correct extracted records, sacrificing more resources to further improve the extraction accuracy and completeness of RAG.

Based on the data extraction results showed in Table 6, the traditional NLP counting representation, ChemDataExtractor, appears to be gradually being replaced by modern LLM tools. However, it is not necessary to completely abandon ChemDataExtractor, as it can also be used in other ways.[105]

For all tools, an improvement would be to add a "de-contextualization" pre-processing step, which replaces pronouns and other elements in the sentences with their original content in advance through semantic analysis by LLMs, thereby reducing the processing burden on the model during the extraction phase.

For RAG models, improved *Embedding Models* and *Inference Models*, as well as through hyperparameter optimisation, can be envisaged. Also, we can further fine-tune LLMs[106, 107] (whether it's the BERT model in BERT-PSIE or the GPT model in ChatExtract or LangChain) using data from the field of Materials Science to enhance the LLM's understanding of domain knowledge.

For RAG implementation, semantic chunking[108-110] or adaptive retrieval queries[111, 112] may also be a possible way. In addition to adding re-ranking to optimize retrieval,[113] there are also many improved RAG frameworks that can further aid in extraction.[114-116]

We also found that LLMs have a limited ability to follow instructions ("lack of focus"), e.g., if the LLM is simultaneously asked to extract data and output it in a given structure, it is likely to do neither well. Our analysis of the *Error Class* in Section 3.4 also illustrates this as tools struggle to get all the details correct simultaneously. The best approach to this is to provide detailed execution steps and simplify the single task.

It is also possible to combine the information extraction with the latest advances in reinforcement learning to mimic reasoning, such as Chain of Thought (CoT).[117-119] This can be easily achieved by directly replacing existing LLMs with LLMs that have reinforcement learning CoT, or by autonomously building workflows with reasoning nodes.

The extraction of data from tables and figures presents a significant challenge, particularly when dealing with complex visual elements. While the focus here is not on figure extraction, advancements in technology offer promising solutions. For example, the use of models equipped with image analysis capabilities, e.g., visual language models (VLMs)[86] or multimodal large language models (MLLMs),[85] to calculate the *Null-Precision* of the figures (assess whether a figure contains the target data), then humans could manually extract information from this narrowed set of relevant figures.

This work also has some limitations. For example, the number of 200 examined articles is too small to support a more diverse comparison, as the sometimes large obtained error bars indicate. Also, the data extraction is only at the numerical level, without including categorical properties (e.g., data type, bandgap type, crystal structure) that are not required to uniquely identify the *Material* within the source's context. The number of tested LLMs is also insufficient to determine the impact of parameter quantity or quantification accuracy on extraction effectiveness.

We will conduct further research, exploring not only the best ways to use the models but also attempting to more specifically adjust the model architecture, continuously trying the latest technologies to obtain a practically usable data extraction tool, and using it to extract a large-scale experimental Materials Science database.

## Conclusions

The use of Natural Language Processing to extract data from publications could greatly benefit the scientific community. Nevertheless, the language and contextual issues associated with the presentation of scientific (and other) data are complex, leading to significant challenges.

Herein, five representative AI tools were applied to extract bandgap information from 200 randomly selected papers. Except for the optimisation of explicit hyperparameters, the models, and the way they were implemented, were not modified to meet the challenges presented by the individual papers from within our unbiased datasets, and so the results obtained may fall below what could be achievable in the future.

In particular, the results obtained for *Precision* were encouraging, with many tools achieving scores of 40–50 %, but *Recall* is identified as a key current issue, with the best tools only achieving 20 %. Tools based on either ChatExtract or LangChain were found to deliver the best results. A feature that has not been previously widely recognised is that fine details of the data presentation, e.g., document origins from arXiv versus publisher versions, can have a significant influence on the quality of the data extraction. From the perspective of users wanting to apply these tools to extract data, it would be common for highlighted papers to be manually overviewed and key results verified. Central to this is the *Null-Precision* of the tools as high *Null-Precision* minimises such intense manual effort. It is encouraging that most tools showed *Null-Precision* values exceeding 94 %.

Nevertheless, the issue of extracting elements located in context challenge all current tools. The development of a comprehensive database of bandgap values would be a major contribution to the advancement of Materials Science, but this must be done with high *Precision*, *Recall*, and *Null-Precision*.

## Author contributions

W.N. designed the research, performed all data extractions, Python programming to establish the combined spreadsheet, performed manual data extractions and data analysis, and wrote the manuscript. M.L. contributed to the research design and implementation. J.R.R. supervised data analysis design and implementation, as well as editing the manuscript. R.K. conceived and supervised the project.

## Conflicts of interest

There are no conflicts to declare.

## Data availability

All mathematical analyses of the data are provided in ESI. All developed source codes, inputs and outputs, plus Excel files depicting results from each stage of the data extraction process, are available at GitHub (https://github.com/wenkaining/Bandgap-Extraction-Comparison), DOI: 10.5281/zenodo.17785333.

## Acknowledgements

We thank the Australian Research Council for funding this research under Grant CE230100021, The Young Scientists Fund of the National Natural Science Foundation of China (Grant No. 12404276), the Special Funds of the National Natural Science Foundation of China (Grant No. 12347164), the China Postdoctoral Science Foundation (Grant Nos. 2024T170541 and GZC20231535), and the University of Technology Sydney for the provision of computing resources at National Computational Infrastructure, Australia.